\documentclass[twocolumn,showpacs,showkeys,amsmath,amssymb,superscriptaddress,floatfix,nofootinbib]{revtex4}

\usepackage{graphicx}
\usepackage{bm} 
\usepackage{subfigure}

\makeatletter
\newcommand\erfc{\mathop{\operator@font erfc}\nolimits}
\def\slashchar#1{\setbox0=\hbox{$#1$}
   \dimen0=\wd0 \setbox1=\hbox{/} \dimen1=\wd1
   \ifdim\dimen0>\dimen1 \rlap{\hbox to \dimen0{\hfil/\hfil}} #1
   \else  \rlap{\hbox to \dimen1{\hfil$#1$\hfil}} / \fi}

\begin{document}
 
\title{
Single-freeze-out model for ultra relativistic heavy-ion collisions at $\sqrt{s_{\rm NN}}=2.76$~TeV
and the LHC proton puzzle}

\author{Maciej Rybczy\'nski} 
\email{Maciej.Rybczynski@ujk.edu.pl}
\affiliation{Institute of Physics, Jan Kochanowski University, PL-25406~Kielce, Poland} 

\author{Wojciech Florkowski} 
\email{Wojciech.Florkowski@ifj.edu.pl}
\affiliation{Institute of Physics, Jan Kochanowski University, PL-25406~Kielce, Poland} 
\affiliation{The H. Niewodnicza\'nski Institute of Nuclear Physics, Polish Academy of Sciences, PL-31342 Krak\'ow, Poland}

\author{Wojciech Broniowski} 
\email{Wojciech.Broniowski@ifj.edu.pl}
\affiliation{Institute of Physics, Jan Kochanowski University, PL-25406~Kielce, Poland} 
\affiliation{The H. Niewodnicza\'nski Institute of Nuclear Physics, Polish Academy of Sciences, PL-31342 Krak\'ow, Poland}

\date{February 24, 2012}

\begin{abstract}
The single-freeze-out model with parametrized hypersurface and flow geometry is employed to analyze the transverse-momentum spectra of hadrons produced in the Pb+Pb collisions at the collision energy of \mbox{$\sqrt{s_{\rm NN}}=2.76$~TeV} at the CERN Large Hadron Collider (LHC). With the notable exception for protons and antiprotons, we find a very good agreement between the model results and the data for the measured hadron species. The additional analysis of the HBT radii of pions helps us to select, from several different types of freeze-out studied in this work, the most realistic form of the freeze-out hypersurface. We find that discrepancy ratio between the model and experiment for the proton/antiproton spectra depends on $p_T$, dropping from 2 in the soft region to 1 around $p_T=1.5$~GeV.
\end{abstract}

\pacs{25.75.-q, 25.75.Dw, 25.75.Ld}

\keywords{relativistic heavy-ion collisions, particle spectra, femtoscopy, LHC}

\maketitle 

\section{Introduction}
\label{sect:intro}

Thermal approach \cite{Koch:1985hk,Cleymans:1992zc,Csorgo:1995bi,Rafelski:1996hf,Rafelski:1997ab,Cleymans:1998fq,Cleymans:1999st,Gazdzicki:1998vd,Gazdzicki:1999ej,Braun-Munzinger:1994xr,Cleymans:1996cd,Becattini:2000jw,Sollfrank:1993wn,Schnedermann:1993ws,Braun-Munzinger:1995bp,Becattini:1997uf,Yen:1998pa,Braun-Munzinger:1999qy,Becattini:2003wp,Braun-Munzinger:2001ip,Florkowski:2001fp,Broniowski:2001we,Broniowski:2001uk,Broniowski:2002nf,Retiere:2003kf} has become one of the cornerstones of our understanding of ultra-relativistic heavy-ion collisions, allowing to explain the data on hadronic abundances collected at the AGS \cite{Braun-Munzinger:1994xr,Cleymans:1996cd,Becattini:2000jw}, SPS \cite{Sollfrank:1993wn,Schnedermann:1993ws,Braun-Munzinger:1995bp,Becattini:1997uf,Yen:1998pa,Braun-Munzinger:1999qy,Becattini:2003wp}, and RHIC energies \cite{Braun-Munzinger:2001ip,Florkowski:2001fp,Broniowski:2001we,Broniowski:2001uk,Broniowski:2002nf,Retiere:2003kf}.

Amended with the single-freeze-out scenario and a proper choice of the freeze-out geometry \cite{Broniowski:2001we,Broniowski:2001uk,Broniowski:2002nf}, it also properly describes the RHIC transverse momentum spectra \cite{Florkowski:2002wn,Baran:2003nm}, collective flow \cite{Retiere:2003kf,Broniowski:2002wp,Florkowski:2004du}, femtoscopic observables \cite{Kisiel:2006is,Broniowski:2008vp,Kisiel:2008ws}, or certain two-particle correlation variables, such as the charge balance functions \cite{Florkowski:2004em,Bozek:2003qi}. One obtains successful fits economically, with just a few thermodynamic parameters, such as the temperature, $T$, the baryon and strangeness chemical potentials, $\mu_B$ and $\mu_S$, or in the extended approach the quark fugacities, $\gamma_s$ and $\gamma_q$, and several parameters describing the geometry and flow. For that reason in many popular approaches the thermal (or statistical) approach is used to model the hadronization stage of the heavy-ion reaction.  

Much to our surprise, the first measurements at the LHC \cite{Kalweit} confirmed the validity of the thermal model for all identified particle species, but for the protons! This {\em proton puzzle} is indeed perplexing, as protons (and antiprotons) are basic products of the reaction and as such they need to be described in any successful approach. In this paper we argue that the puzzle is in some sense even deeper, as not only the abundances of all the other measured particles are correctly reproduced, but also their transverse-momentum spectra can be described without difficulty. Thus, it is solely the protons (and antiprotons), where the model fails to describe these basic observables. 

In the present study we shall apply simple parameterizations of the freeze-out geometry, similarly as in our early work \cite{Broniowski:2001we,Broniowski:2001uk,Broniowski:2002nf}, which are motivated by the dynamical studies involving relativistic hydrodynamics \cite{Teaney:2001av,Hirano:2002ds,Kolb:2003dz,Huovinen:2003fa,Shuryak:2004cy,Eskola:2005ue,Hama:2005dz,Hirano:2005xf,Huovinen:2006jp,Hirano:2007xd,Nonaka:2006yn,Broniowski:2008vp,Huovinen:2009yb}. In particular, we will explore the Cracow \cite{Broniowski:2001we} and the Blast-Wave models \cite{Danielewicz:1992mi,Schnedermann:1993ws,Retiere:2003kf,Florkowski:2004tn}.
The advantage of such an approach is that one can focus entirely on the freeze-out and flow geometry, avoiding 
intricacies of the earlier dynamical evolution. 

Our basic result is that the identified transverse-momentum spectra of all up-to-now measured particles at the LHC, {\em with the important exception of protons}, can be properly fitted, describing the Pb+Pb data at \mbox{$\sqrt{s_{\rm NN}}=2.76$~TeV} in the soft regime. We also test the reliability of the fits with the HBT correlation data, which allows us to select the most realistic freeze-out conditions out of several different types of freeze-out studied in this work.

\section{Method}

As mentioned above, the concept of the single freeze-out \cite{Broniowski:2001we,Broniowski:2001uk,Broniowski:2002nf,Blume:2011kq} allows for the uniform calculations of numerous soft hadronic observables. The simulations presented in this work have been carried out with {\tt THERMINATOR} \cite{Kisiel:2005hn,Chojnacki:2011hb}.  Monte-Carlo method allows for a simple inclusion of the experimental cuts and, therefore, for more realistic verification of the model predictions.  

At the extreme energies studied at the LHC, we expect that the midrapidity region contains equal numbers of baryons and antibaryons, hence the values of the chemical potentials used in this work are set equal to zero ($\mu_B=\mu_S=\mu_I=0$, where $\mu_I$ is the chemical potential related to the isospin conservation).  Hence,  we are left with temperature, $T$, as the only independent thermodynamic parameter. A natural expectation, based on the shape of the $T-\mu_B$ freeze-out curve \cite{Cleymans:1998fq,Cleymans:1999st} and confirmed by the analysis of the hadronic abundances at the LHC \cite{Kalweit}, is that the temperature of the chemical freeze-out at the LHC should be very similar to the value determined at RHIC. Therefore, we use $T_{\rm chem} = 165.6$~MeV \cite{Chojnacki:2011hb}. We note that this value is close to the value of the transition temperature of the crossover found at $\mu_B=0$ in the lattice simulations of QCD \cite{Borsanyi:2011bn}.

Having fixed the values of thermodynamic parameters, we test different parameterizations for the shape of the freeze-out hypersurface and for the form of flow at freezeout to achieve the best description of the transverse-momentum spectra of charged pions and kaons. We stress that in our method {only these two hadron species are used to determine the geometric and expansion parameters}, since they are experimentally determined with the best accuracy. We fit the Cracow and Blast-Wave models (by the least squares method) to the experiment. Then, the spectra of other hadrons are predictions of our approach, allowing for its verification.

\begin{figure}[b] 
\begin{center}
\includegraphics[angle=0,width=0.3\textwidth]{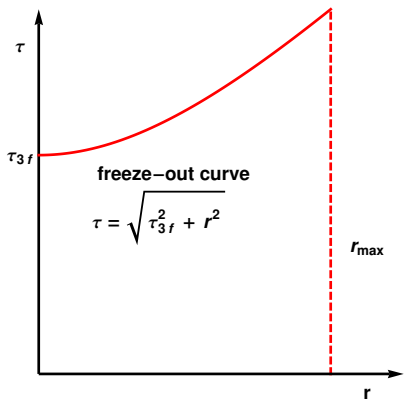}
\end{center}
\caption{Projection of the freeze-out hypersurface used in the Cracow model on the $\tau-r$ plane (\mbox{$\tau=\sqrt{t^2-z^2}$} and \mbox{$r=\sqrt{x^2+y^2}$}). The parameters $\tau_{\rm 3f \,}$ and $r_{\rm max}$ are the two geometric parameters of the model. }
\label{fig:hubble}
\end{figure} 

In the case of the Cracow model, we fit the value of the proper time at freeze-out and the transverse size of the fire-cylinder. In the case of the generalized Blast-Wave model, we fit the proper time and the magnitude of the transverse flow for three different values of the slope of the freeze-out curve in Minkowski space, controlled by a parameter $A$. The three different choices of $A$ correspond to three different physical scenarios: if $A$ is positive the freeze-out starts at the center of the system, if $A$ is negative the freeze-out starts at the edges, and for $A=0$ the freeze-out happens at constant (longitudinal proper) time in the whole volume.   

With the procedure outlined above, we find a very good agreement between the model results and the experimental data for all measured hadron species excluding protons. Moreover, with the hadron spectra alone, no preference for any of the employed freeze-out models can be found. A further analysis of the pion HBT radii indicates, however, that the freeze-out 
hypersurface used in the generalized Blast-Wave model with $A=-0.5$ leads to the best agreement between the data and theory. This confirms an earlier observation done for RHIC \cite{Kisiel:2006is}. This type of freeze-out 
is also consistent with the hydrodynamic picture where the freeze-out starts at the edges of the system and continues inwards. 

\section{Cracow model}
\label{sect:model-crac}

\subsection{Definition of freeze-out conditions}
\label{sect:def-crac}

The Cracow model assumes boost-invariant and cylindrically symmetric conditions at freeze-out. The freeze-out hypersurface is defined by the condition \cite{Broniowski:2001we}
\begin{equation}
t^2 - x^2 - y^2 - z^2 = \tau_{\rm 3f \,}^2 = \hbox{const.},
\label{fh-crac}
\end{equation}
and the fluid four-velocity is proportional to the space-time position
\begin{equation}
u^\mu = \gamma ( 1, {\bf v}) = {x^\mu \over \tau_{\rm 3f \,}}={t \over \tau_{\rm 3f \,}}
\left(1,{ {\bf x} \over t} \right).
\label{umu-crac}
\end{equation}
The index $3$ in the symbol $\tau_{\rm 3f \,}$ reminds us that the freeze-out takes place on the three-dimensional hypersurface of constant proper time, see Fig.~\ref{fig:hubble}. 

Equations (\ref{fh-crac}) and (\ref{umu-crac}) imply that the parameterizations of the freeze-out hypersurface and the four-velocity field are connected. The longitudinal proper time, \mbox{$\tau=\sqrt{t^2-z^2}$}, and the transverse distance, \mbox{$r=\sqrt{x^2+y^2}$}, may be expressed in terms of the transverse rapidity $\eta_\perp$,
\begin{eqnarray}
\tau &=&  \tau_{\rm 3f \,} \, \cosh \eta_\perp, \nonumber \\
r &=& \tau_{\rm 3f \,} \, \sinh \eta_\perp .
\label{eta_perp}
\end{eqnarray}
We also use the parameterizations \mbox{$t =  \tau  \cosh \eta_\parallel$} and \mbox{$z = \tau  \sinh \eta_\parallel$} where \mbox{$\eta_\parallel = 1/2 \, \ln(t-z)/(t-z)$} is the space-time rapidity, and also \mbox{$x =  r  \cos \phi$} and \mbox{$y = r  \sin \phi$} where $\phi$ is the azimuthal angle. 

The calculation of the volume element of the freeze-out hypersurface in the Cracow model shows that it is proportional to the four-velocity (as in the original Blast-Wave model by Siemens and Rasmussen \cite{Siemens:1978pb}), 
\begin{eqnarray}
d\Sigma^\mu &=& u^\mu \tau_{\rm 3f \,}^3 \hbox{cosh}(\eta_\perp)
\hbox{sinh}(\eta_\perp) \,d\eta_\perp \,d\eta_\parallel \,d\phi \nonumber \\
&=& u^\mu \tau_{\rm 3f \,} r \,dr \,d\eta_\parallel \,d\phi.
\label{dS-crac}
\end{eqnarray}
The formula (\ref{dS-crac}) is used in the Cooper-Frye formula,  
\begin{equation}
\frac{dN}{dy dp_\perp} = 2 \pi p_\perp \int d\Sigma_\mu(x) p^\mu f(u_\nu p^\nu),
\label{cp1}
\end{equation}
to generate hadronic states on the freeze-out hypersurface defined by the condition (\ref{fh-crac}). In Eq. (\ref{cp1}), the function $f(u_\nu p^\nu)$ is the distribution function which is taken in the form of the Bose-Einstein or Fermi-Dirac distribution for bosons and fermions, respectively, and $p^\mu$ is the particle's four-momentum expressed by the rapidity $y$ and the transverse-momentum $p_\perp$ (the quantity $m_\perp = \sqrt{m^2 + p_\perp^2}$ is the transverse mass),
\begin{equation}
p^\mu = \left( m_\perp \cosh y, p_\perp \cos \phi_p,  p_\perp \sin \phi_p,  m_\perp \sinh y \right).
\end{equation}

The generation of particles is performed  with {\tt THERMINATOR~2} \cite{Chojnacki:2011hb} which includes all known hadrons containing $u,d$ and $s$ quarks ({\tt THERMINATOR~2} has the same particle input basis as {\tt SHARE} \cite{Torrieri:2004zz}). {\tt THERMINATOR~2} simulates also decays of resonances, hence, the final spectra consist of primordial and secondary contributions (the primordial particles are emitted directly from the fireball, while the secondary particles come from the resonance decays). 

\begin{figure}[t] 
\begin{center}
\includegraphics[angle=0,width=0.4\textwidth]{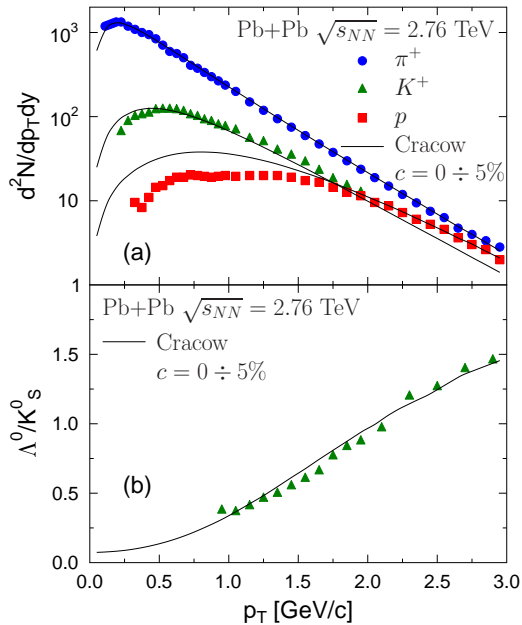}
\end{center}
\caption{\small {\bf (a)} Transverse-momentum spectra of pions (circles), kaons (triangles), and protons (squares) for the centrality class $c=0-5$\% \cite{Preghenella}. The Cracow model results (solid lines) generated with \texttt{THERMINATOR~2} \cite{Chojnacki:2011hb} were obtained for: $T=165.6$~MeV, $\mu_B=0$, $\tau_{\rm 3f} =9.0$~fm, and $r_{\rm max}=11.4$~fm. The geometric parameters were fitted to the spectra of pions and kaons, only. {\bf (b)} The model ratio of the transverse-momentum spectra of $\Lambda^0$'s and $K^0_S$'s (solid line), compared with the data (triangles) for the same centrality class $c=0-5$\% \cite{Kalinak}. The values of the thermodynamic and geometric parameters are the same in the two parts. }
\label{fig:mostcentral-cracow}
\end{figure} 

The Cracow single-freeze-out model has altogether four parameters: two geometric and two thermodynamic ones. The two geometric parameters are $\tau_{\rm 3f \,}$ and $r_{\rm max}$, while the two thermodynamic parameters are temperature, $T$, and baryon chemical potential, $\mu_B$. The thermodynamic parameters define the local equilibrium distribution functions $f(u_\nu p^\nu)$ in Eq. (\ref{cp1}). 

Since the values of $T$ and $\mu_B$ follow from the analyses of hadron abundances, in the present work we may use the results of the previous studies where the conditions for chemical equilibrium at the LHC have been studied, \cite{Andronic:2007vh}. On more general grounds, in the central region of the heavy-ion collisions performed at the LHC energy, we expect equal numbers of baryons and antibaryons which implies $\mu_B=0$ (the other chemical potential should also vanish). This leaves us with temperature as the only independent thermodynamic parameter. Expecting that the freeze-out temperature is the same as that found for RHIC \cite{Florkowski:2001fp} we use the value, $T = 165.6$ MeV.

\subsection{Comparison with the LHC data}
\label{sect:lhc-crac}


In Fig.~\ref{fig:mostcentral-cracow}~(a) we show the experimental transverse-momentum spectra of pions (circles), kaons (triangles), and protons (squares) \cite{Preghenella} for the centrality class \mbox{$c=0-5$\%}~\footnote{All the data and the model fits refer to Pb+Pb collisions at $\sqrt{s_{\rm NN}} =2.76$ TeV. Moreover, to be consistent with the experimental procedure \cite{Kalinak}, the spectra of protons are corrected for the decays of $\Lambda^0$'s and $\Sigma^0$'s. Similarly, $\Lambda^0$'s are corrected for the decays of $\Xi^-$'s. }. The Cracow model results generated with \texttt{THERMINATOR~2} \cite{Chojnacki:2011hb} are denoted by the solid lines. The parameters used in the model calculations are: \mbox{$T=165.6$~MeV}, $\mu_B=0$, $\tau_{\rm 3f} =9.0$~fm, and $r_{\rm max}=11.4$~fm. The values of the thermodynamic parameters have been treated as external parameters and the geometric parameters were fitted (with the least squares method) to the spectra of pions and kaons, only. 

In Fig.~\ref{fig:mostcentral-cracow}~(b) we show the model ratio of the transverse-momentum spectra of $\Lambda^0$'s and $K^0_S$'s (solid line), compared with the data (triangles) for the same centrality class $c=0-5$\% \cite{Kalinak}. The values of the thermodynamic and geometric parameters are the same in the two parts of Fig.~\ref{fig:mostcentral-cracow}. 

We observe very good agreement between the data and the model results for pions, kaons, and $\Lambda^0$'s in the soft momentum region, \mbox{$p_\perp \leq$ 3 GeV}. On the other hand, the model results for protons overpredict the data. This is an expected result, since already the results for the abundances have indicated that the thermal models used in the grand-canonical version cannot predict correctly the ratio of pion and proton abundances \cite{Kalweit}. 

\begin{figure}[t] 
\begin{center}
\includegraphics[angle=0,width=0.4\textwidth]{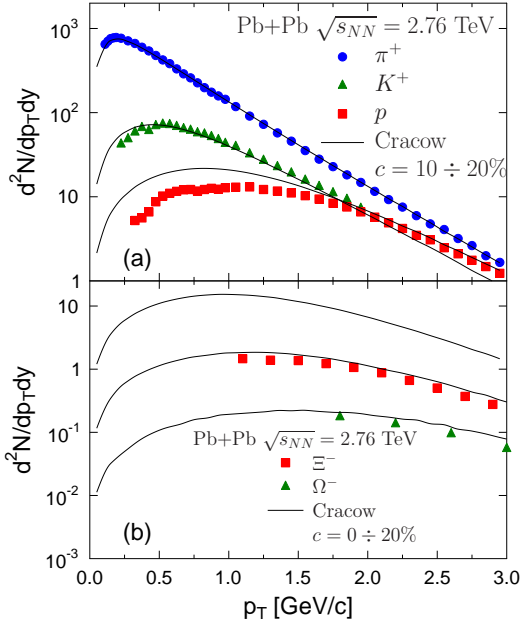}
\end{center}
\caption{\small {\bf (a)} Transverse-momentum spectra of pions (circles), kaons (triangles), and protons (squares) for the centrality class $c=10-20$\% \cite{Preghenella}. The Cracow model parameters used in the calculation: $T=165.6$~MeV, $\mu_B=0$, $\tau_{\rm 3f} =7.4$~fm, and $r_{\rm max}=9.6$~fm. Again, the geometric parameters were fitted to the spectra of pions and kaons, only. {\bf (b)} The model transverse-momentum spectra of $\Lambda^0$'s, $\Xi^-$'s, and $\Omega^-$'s, compared with the experimental results for $\Xi^-$'s and $\Omega^-$'s for the centrality class $c=0-20$\% \cite{Kalweit}. The values of the thermodynamic and geometric parameters are the same in the two parts. Since the geometric parameters were fitted to the centrality class $c=10-20$\%, the model results for hyperons are slightly above the experimental data. }
\label{fig:semicentral-cracow}
\end{figure} 


In Fig.~\ref{fig:semicentral-cracow} (a) we show again the transverse-momentum spectra of pions (circles), kaons (triangles), and protons (squares) but now for the centrality class $c=10-20$\% \cite{Preghenella}. The geometric parameters used in the model calculation are: $\tau_{\rm 3f} =7.4$~fm, and $r_{\rm max}=9.6$~fm. Similarly to the case of central collisions, the geometric parameters were fitted to the spectra of pions and kaons, only. In the natural way, they are smaller than those found in the case of the most central collisions. 

In Fig.~\ref{fig:semicentral-cracow} (b) we show the model transverse-momentum spectra of $\Lambda^0$'s, $\Xi^-$'s, and $\Omega^-$'s. They are compared with the experimental results for $\Xi^-$'s (squares) and $\Omega^-$'s (triangles) for the centrality class $c=0-20$\% \cite{Kalinak}. The values of the thermodynamic and geometric parameters are the same in the parts (a) and (b). Since the geometric parameters were fitted to the centrality class $c=10-20$\%, the model results for hyperons are slightly above the experimental data but their $p_\perp$ dependence is consistent with the data. 

Similarly to the central collisions, we see again that the agreement between the data and the model predictions is very good except for the protons. 

\begin{figure}[t]  
\begin{center}
\includegraphics[angle=0,width=0.4\textwidth]{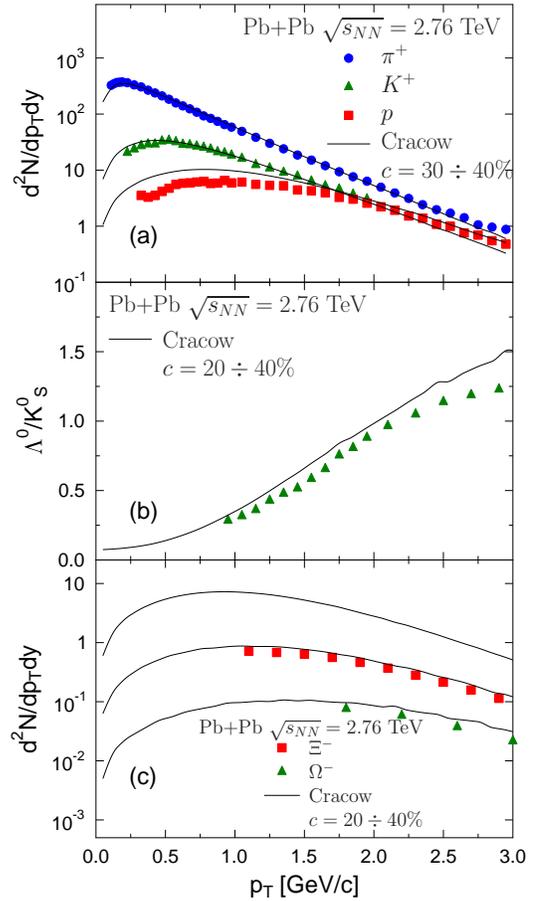}
\end{center}
\caption{\small {\bf (a)} Transverse-momentum spectra of pions (circles), kaons (triangles), and protons (squares) for the centrality class $c=30-40$\% \cite{Preghenella}. The Cracow model parameters used in the calculation: $T=165.6$~MeV, $\mu_B=0$, $\tau_{\rm 3f} =5.9$~fm, and $r_{\rm max}=7.25$~fm. Once again, the geometric parameters were fitted to the spectra of pions and kaons, only. {\bf (b)} The model ratio of the transverse-momentum spectra of $\Lambda^0$'s and $K^0_S$'s, compared to the data for the centrality class $c=20-40$\% \cite{Kalinak}. {\bf (c)} The model transverse-momentum spectra of $\Lambda^0$'s, $\Xi^-$'s, and $\Omega^-$'s, compared with the experimental results for $\Xi^-$'s and $\Omega^-$'s for the centrality class $c=20-40$\% \cite{Kalweit}. The values of the thermodynamic and geometric parameters are the same in the three parts. }
\label{fig:lesscentral-cracow}
\end{figure}  


In Fig.~\ref{fig:lesscentral-cracow} (a) we show once again the transverse-momentum spectra of pions (circles), kaons (triangles), and protons (squares) but this time for the centrality class $c=30-40$\% \cite{Preghenella}. The Cracow model parameters used in the calculation are: $\tau_{\rm 3f} =5.9$~fm, and $r_{\rm max}=7.25$~fm. They were again fitted to the spectra of pions and kaons, only. The decreasing trend of $\tau_{\rm 3f}$ and $r_{\rm max}$ with increasing centrality reflects the decrease of multiplicity.

Figure \ref{fig:lesscentral-cracow} (b) shows the model ratio of the transverse-momentum spectra of $\Lambda^0$'s and $K^0_S$'s, compared to the data (triangles) for the centrality class \mbox{$c=20-40$\%} \cite{Kalinak}, whereas Fig.~\ref{fig:lesscentral-cracow} (b) shows the model transverse-momentum spectra of $\Lambda^0$'s, $\Xi^-$'s, and $\Omega^-$'s, compared with the experimental results for $\Xi^-$'s (squares) and $\Omega^-$'s (triangles) for the centrality class $c=20-40$\% \cite{Kalweit}. The values of the thermodynamic and geometric parameters are the same in the three parts. Similarly to Fig.~\ref{fig:semicentral-cracow}, the model hyperon spectra are slightly above the data, which may be explained by different centrality classes analyzed in the presented comparison.

\begin{figure}[b]
\begin{center}
\includegraphics[angle=0,width=0.3\textwidth]{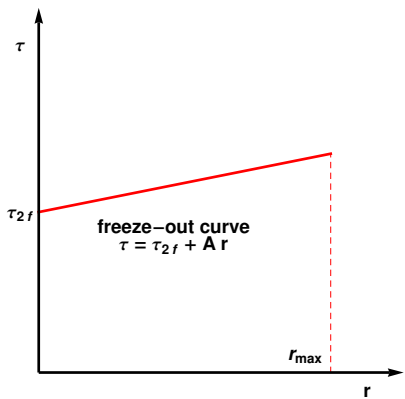}
\end{center}
\caption{The freeze-out curves considered in the modified Blast-Wave model \cite{Kisiel:2006is}.}
\label{fig:model-a}
\end{figure}

Our comparison of the transverse-momentum spectra obtained within the Cracow model with the experimental data may be summarized with the statement that the model describes well the spectra of all measured hadrons except for protons, in the $p_\perp$ range up to 3 GeV. 

\section{Generalized Blast-Wave Model}
\label{sect:model-bwa}

\subsection{Definition of freeze-out conditions}
\label{sect:def-bwa}

Probably, the most popular parameterization of the freeze-out hypersurface is the Blast-Wave model 
\cite{Schnedermann:1993ws,Retiere:2003kf}. In its standard form, the model is boost-invariant and cylindrically symmetric (similarly as the Cracow model discussed in the previous Section). The Blast-Wave model uses the assumption that the freeze-out happens at a constant value of the longitudinal proper time
\begin{equation}
\tau = \sqrt{t^2 - z^2} = \tau_{\rm 2f} = \hbox{const}.
\label{fh-a-1}
\end{equation}
In order to get a broader applicability, we generalize this condition to the formula
\begin{equation}
\tau = \tau_{\rm 2f}  + A \, r,
\label{fh-a-2}
\end{equation}
where $\tau_{\rm 2f}$ and $A$ are constants, and $A$ describes the slope of the freeze-out curve in the Minkowski space, see Fig.~\ref{fig:model-a}. With $A>0$ ($A<0$) we may consider the freeze-out scenarios where the outer parts of the system freeze-out later (earlier). Of course, with $A=0$ we reproduce the standard Blast-Wave parametrization (\ref{fh-a-1}).

In the generalized Blast-Wave model we find compact expressions for the argument of the equilibrium distributions functions,
\begin{eqnarray}
u_\nu p^\nu &=& 
\frac{\left[m_\perp \cosh(\eta_\parallel - y) - {\tilde v}_\perp(r)\, p_\perp \cos(\phi - \phi_p)\right]}
{\sqrt{1-{\tilde v}_\perp^2(r)}} \nonumber \\
\label{A-pu}
\end{eqnarray}
and for the Cooper-Frye integration measure,
\begin{eqnarray}
d\Sigma_\mu p^\mu 	&=& (\tau_{\rm 2f} + A\, r)\, r\, \left[m_\perp \cosh(\eta_\parallel - y) \right. \nonumber \\
&& \left.  - A\, p_\perp \cos(\phi - \phi_p) \right]. 
\label{A-Sp}
\end{eqnarray}
Equations (\ref{A-pu}) and (\ref{A-Sp}) should be used in the Cooper-Frye formula (\ref{cp1}).

The user of {\tt THERMINATOR~2} \cite{Chojnacki:2011hb} may choose different $r$-profiles of the transverse flow ${\tilde v}_\perp(r)$. In this work we use the following option 
\begin{equation}
\tilde v_\perp(r) = \frac{r/r_{\rm max}}{v_T + r/r_{\rm max}},
\end{equation}
where $v_T$ is the parameter controlling the strength of the transverse flow. In this version of the Blast-Wave model, we have four parameters: $A$, $\tau_{\rm 2f}$, $r_{\rm max}$, and $v_T$.

\begin{table}[t]
{\begin{tabular}{cccc} \toprule
    &           &                    &           \\ 
$A$ & $c\, [\%]$  & $r_{\rm max}$ [fm] & $v_T$     \\ 
    &           &                    &           \\  \colrule
    &           &                    &           \\ 
0.5 & 0--5      & 9.9                &  0.375    \\ 
0.5 & 10--20    & 8.3                &  0.375    \\ 
0.5 & 30--40    & 6.5                &  0.425    \\ 
    &           &                    &           \\  \colrule
    &           &                    &           \\ 
0.0 & 0--5      & 9.9                &  0.45    \\ 
0.0 & 10--20    & 8.2                &  0.43    \\ 
0.0 & 30--40    & 6.2                &  0.46    \\ 
    &           &                    &           \\  \colrule
    &           &                    &           \\ 
-0.5 & 0--5      & 10.4               &  0.46    \\ 
-0.5 & 10--20    & 8.9                &  0.475    \\ 
-0.5 & 30--40    & 6.9                &  0.58    \\  
    &           &                    &           \\  \botrule
\end{tabular}
}
\caption{Optimum choices of the parameters $r_{\rm max}$ and $v_T$ for three fixed values of the parameter $A$ and for different experimental centrality classes. The fits have been performed to the transverse-momentum spectra of pions and kaons only. The fit has been constrained by the condition $\tau_{\rm 2f}/r_{\rm max}=1$. }
\label{tab:modelapar}
\end{table}

\begin{figure}[t] 
\begin{center}
\includegraphics[angle=0,width=0.4\textwidth]{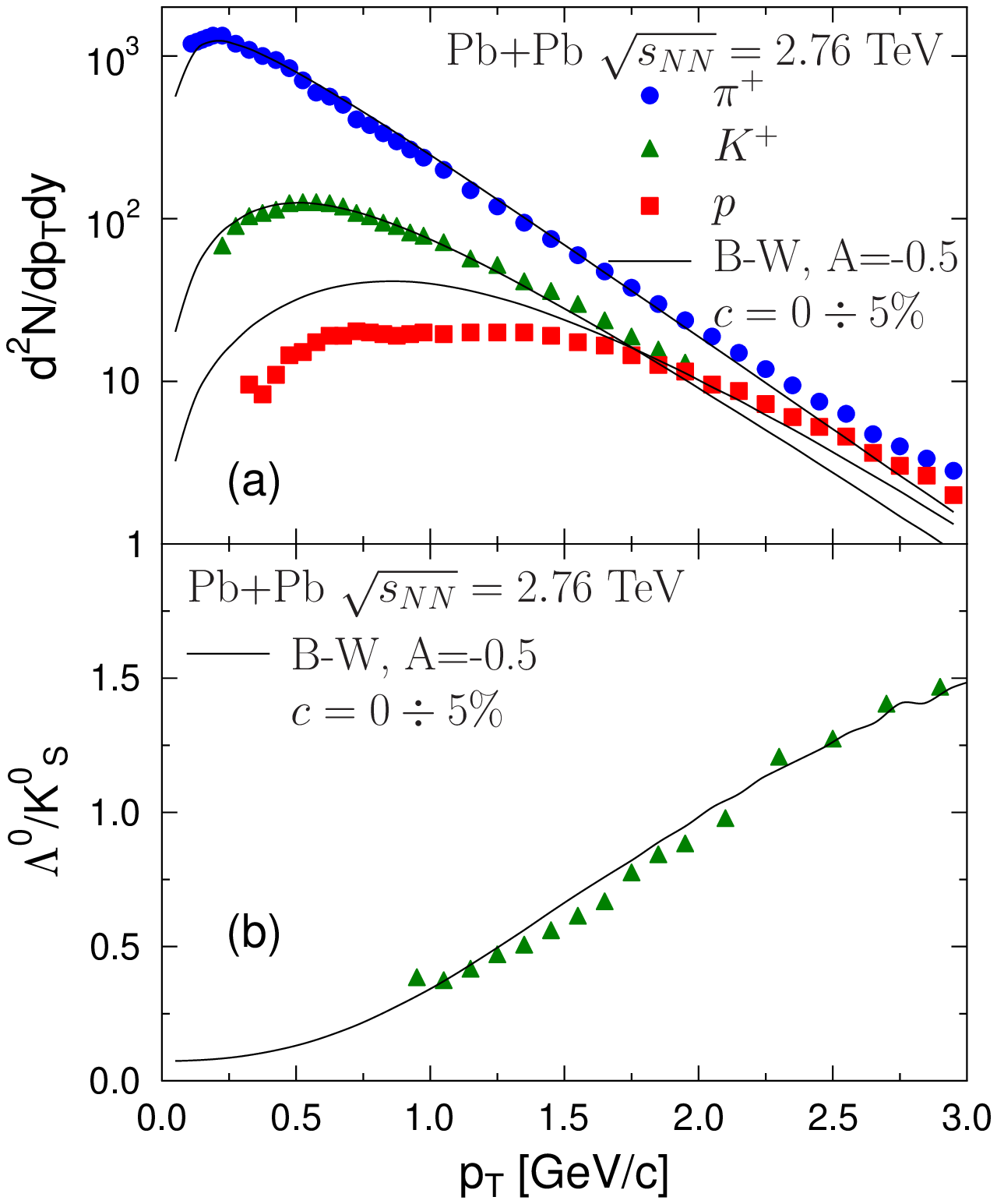}
\end{center}
\caption{\small Same as Fig.~\ref{fig:mostcentral-cracow} but the data are compared to the results obtain with
the Blast-Wave model with $A=-0.5$. The data are taken from \cite{Preghenella,Kalinak}.}
\label{fig:mostcentral-bwa_n}
\end{figure} 

\begin{figure}[t] 
\begin{center}
\includegraphics[angle=0,width=0.4\textwidth]{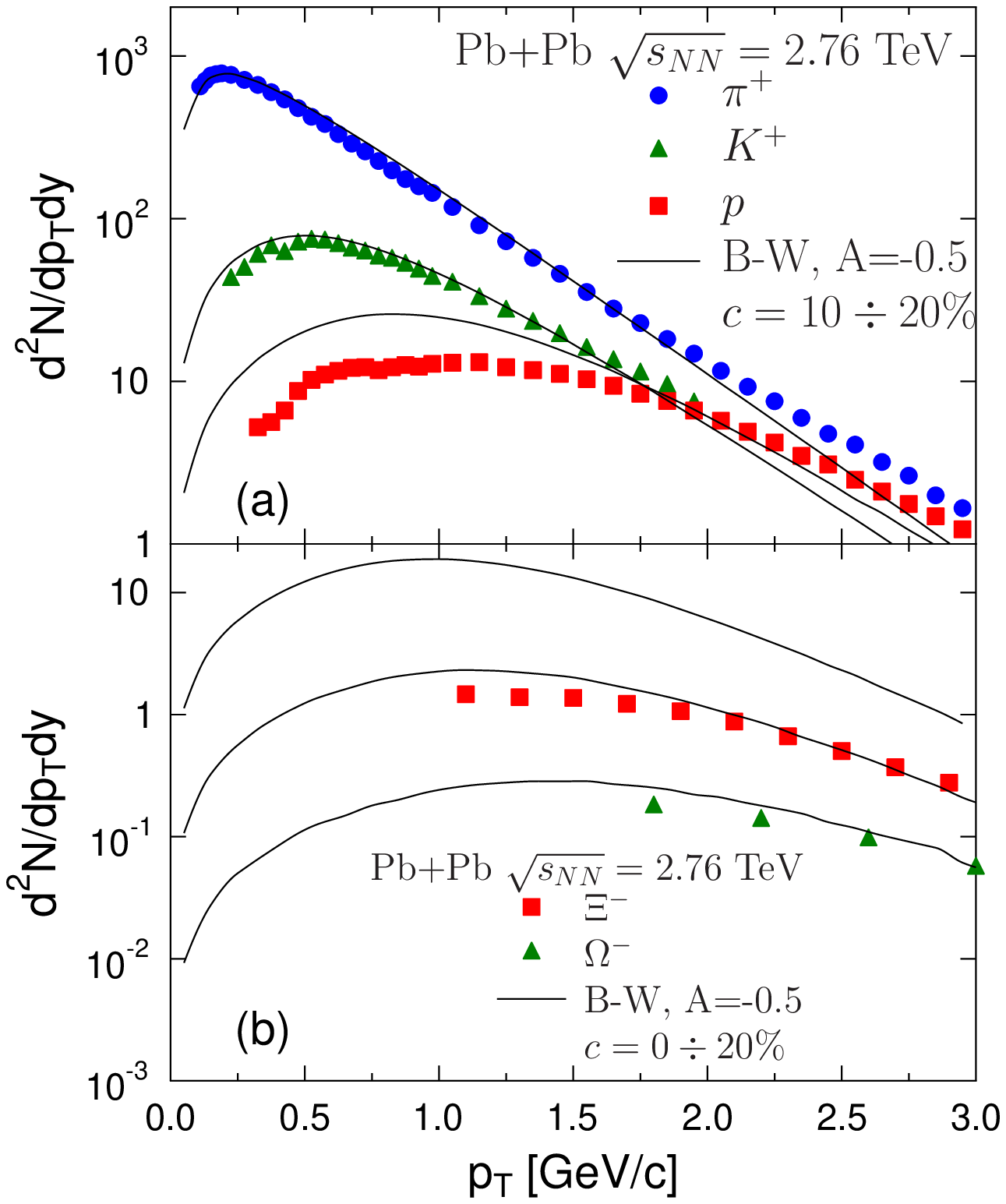}
\end{center}
\caption{\small Same as Fig.~\ref{fig:semicentral-cracow} but the data are compared to the results obtain with
the Blast-Wave model with $A=-0.5$. The data are taken from \cite{Preghenella,Kalweit}.}
\label{fig:semicentral-bwa_n}
\end{figure} 

\begin{figure}[t]  
\begin{center}
\includegraphics[angle=0,width=0.4\textwidth]{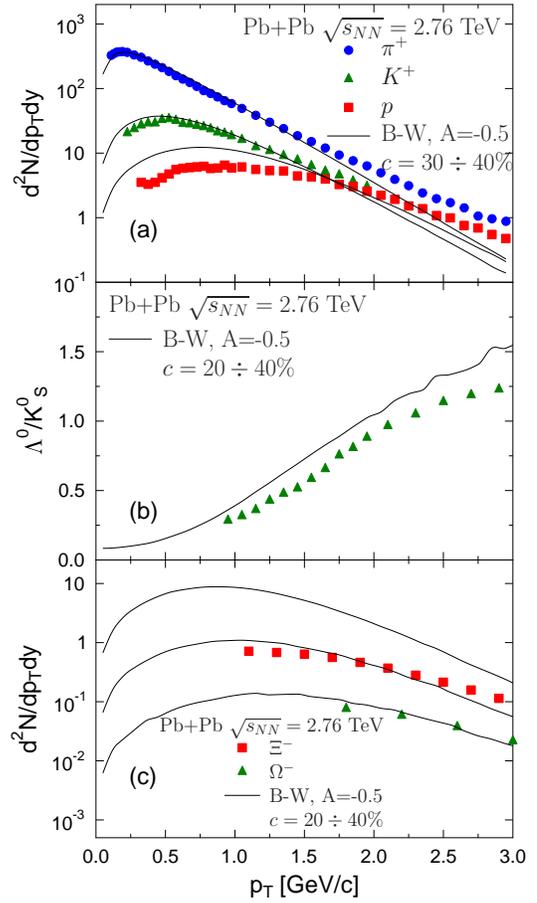}
\end{center}
\caption{\small Same as Fig.~\ref{fig:lesscentral-cracow} but the data are compared to the results obtain with
the Blast-Wave model with $A=-0.5$. The data are taken from \cite{Preghenella,Kalinak,Kalweit}.}
\label{fig:lesscentral-bwa_n}
\end{figure}  

\subsection{Comparison with the LHC data}
\label{sect:lhc-bwa}

The freeze-out conditions defined by the generalized Blast-Wave model with different values of $A$ were studied in Ref.~\cite{ Kisiel:2006is}. One of the conclusions of this work for RHIC is that the transverse-momentum spectra of hadrons can be well reproduced for a wide range of the parameters $A$, however, the HBT radii are quite sensitive to the space-time profile of the freeze-out hypersurface, i.e., to the specific choice of the parameter $A$. 

Inspired by Ref.~\cite{ Kisiel:2006is}, we have performed the analysis of the LHC data in the similar way. At first, we have chosen three values of $A$ ($A=0.5,0,-0.5$) and for each of these values we have found the optimal geometric parameters. In order to reduce the number of independent parameters we have fixed the ratio $\tau_{\rm 2f}/r_{\rm max}$ to unity. In the next step, for each value of $A$ we used the optimal choice of $r_{\rm max}$ and $v_T$ to calculate the HBT radii. 

The optimal choices of the parameters $r_{\rm max}$ and $v_T$ for three fixed values of the parameter $A$ and for different experimental centrality classes are given in Table~\ref{tab:modelapar}. We have found, as suggested by Ref.~\cite{ Kisiel:2006is}, that the transverse-momentum spectra are equally well described for different choices of $A$, if the other parameters are properly chosen. 

In Figs.~\ref{fig:mostcentral-bwa_n}--\ref{fig:lesscentral-bwa_n} we show our fits done with $A=-0.5$. The three figures correspond to  Figs.~\ref{fig:mostcentral-cracow}--\ref{fig:lesscentral-cracow} presented earlier in the context of the Cracow model. We observe again rather good agreement between the model predictions and the data. Noticeable discrepancies can be observed in the ratio of the transverse-momentum spectra of $\Lambda^0$'s and $K^0_S$'s for the centrality class $c=20-40$\%, see Fig.~\ref{fig:lesscentral-bwa_n}~(b). However, these differences are the largest (about 20\%) for the momenta reaching 3 GeV, i.e., in the region where we expect the thermal approach to break down. Moreover, since the thermal approach uses many simplifying assumptions, the agreement within 20\% is usually regarded as quite satisfactory.

\begin{figure}[t]
\begin{center}
\includegraphics[angle=0,width=0.35\textwidth]{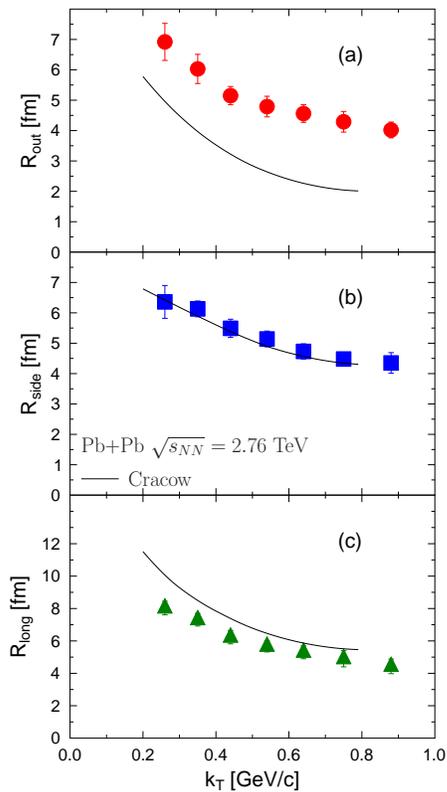}
\end{center}
\caption{The pion HBT radii for the most central collisions obtained in the Cracow model (solid lines) and compared to the LHC data from \cite{Aamodt:2011mr}. The parameters are the same as in Fig.~\ref{fig:mostcentral-cracow}.  
\label{fig:hbtradii_krakow}}
\end{figure}

We do not show here our results obtained for the cases $A=0.5$ and $A=0$. They are very similar to those presented in Figs.~\ref{fig:mostcentral-cracow}--\ref{fig:lesscentral-cracow} and \ref{fig:mostcentral-bwa_n}--\ref{fig:lesscentral-bwa_n}. We note that the shape of the freeze-out hypersurface in the Cracow model is similar to that used in the generalized Blast-Wave model with $A=0.5$ (along the freeze-out curve the proper time grows with the distance from the center). This suggests that the results obtained with the two models should be similar and the values of the optimal close to each other. Indeed, the proper time used in the Blast-Wave model, $\tau_{\rm 2f \,}$, is to a good approximation an average of the proper time and the transverse size used in the Cracow model, $\tau_{\rm 2f \,} \approx  (\tau_{\rm 3f \,}+r_{\rm max})/2$.  

\begin{figure}[t]
\begin{center}
\includegraphics[angle=0,width=0.35\textwidth]{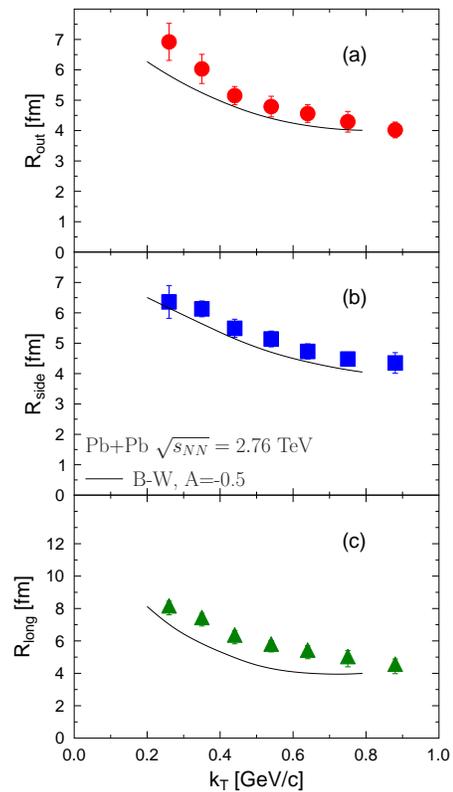}
\end{center}
\caption{The pion HBT radii for the most central collisions obtained in the generalized Blast-Wave model with $A=-0.5$ (solid lines) and compared to the LHC data from \cite{Aamodt:2011mr}. The parameters are the same as in Fig.~\ref{fig:mostcentral-bwa_n}. 
\label{fig:hbtradii_bwan}}
\end{figure}

\section{HBT radii}
\label{sect:lhc-crac-bwa}

Our results presented in the previous Sections indicate that, except for the protons, the transverse-momentum spectra of different hadronic species may be well reproduced in the thermal approach. Moreover, different freeze-out conditions may lead to very similar spectra, as it has been demonstrated by our results obtained with the two different versions of the thermal model. In this Section we present calculations of the HBT radii done in the Cracow model and the Blast-Wave model with $A=-0.5$. It turns out that the use of an additional observable may, in the considered cases, select the most appropriate version of the freeze-out model. 

Figure \ref{fig:hbtradii_krakow} shows the pion HBT radii for the most central Pb+Pb collisions, \mbox{$c=0-5$\%}. Theoretical results obtained in the Cracow model (solid lines) are compared with the LHC data taken from Ref.~\cite{Aamodt:2011mr} ($R_{\rm out}$ -- circles in part (a), $R_{\rm side}$ -- squares in part (b), and $R_{\rm long}$ -- triangles in part (c)). The radii are presented as functions of the transverse-momentum of the pion pair. They have been calculated in {\tt THERMINATOR~2} with the help of the two-particle method (without Coulomb corrections). The parameters used in the simulations are the same as those used to obtain the spectra shown in Fig.~\ref{fig:mostcentral-cracow}. We find that $R_{\rm side}$ is well described, $R_{\rm long}$ is a bit to large, while $R_{\rm out}$ is clearly underpredicted. 

In Fig.~\ref{fig:hbtradii_bwan} we show analogous results for the Blast-Wave model with $A=-0.5$. With much improved agreement of $R_{\rm out}$ with the data, the general consistency between the model calculations and the data has been significantly improved.

\section{Conclusions}
\label{sect:concl}

The single-freeze-out model with parametrized freeze-out hypersurfaces and flow  
has been used to analyze the transverse-momentum spectra of hadrons produced in Pb+Pb collisions 
at the collision energy of \mbox{$\sqrt{s_{\rm NN}}=2.76$~TeV} at the LHC. {\em Except for protons}, we find 
a proper agreement between the model results and the data for all measured hadron species. The additional analysis of the HBT radii of pions suggests that the realistic freeze-out conditions should correspond to an earlier freeze-out of the edges of the system, as suggested by a typical form of the hydrodynamic expansion. 

Thus, the LHC proton puzzle is a certain sense deeper: it appears not only in the simple thermal approach where 
the abundances are calculated, but also in an extension where the $p_T$-spectra can be computed. 
Of course, with the mismatch on abundances, which is a $p_T$-integrated measure, 
we need to find a mismatch in the proton/antiproton spectra. We note, however, from 
Figs.~\ref{fig:mostcentral-cracow} and \ref{fig:mostcentral-bwa_n} that the model proton/antiproton 
spectra are above the data only in the soft region below $p_T \sim 1.5$~GeV, while the harder part is in agreement. 
Therefore the proton puzzle is clearly related to the soft physics. We also note that the ratio of the 
model to experiment is 
$p_T$-dependent, dropping from a value of about 2 at the low-$p_T$ values to $1$ at $p_T \sim 1.5$~GeV. 
Thus a simple rescaling of the proton/antiproton spectra, if found in some treatment, would not do the job. 
This issue, crucial for the thermal approach to relativistic heavy-ion collisions, requires a further study.

\begin{acknowledgements}

This work was supported by the Polish Ministry of Science and Higher Education under Grants No. N N202 263438, N N202 288638, and National Science Centre, grant DEC-2011/01/D/ST2/00772.                                      

\end{acknowledgements}


\end{document}